\documentclass{INTERSPEECH2023}
\usepackage{cite}
\usepackage{amsmath,amssymb,amsfonts}
\usepackage{algorithmic}
\usepackage{graphicx}
\usepackage{textcomp}
\usepackage{multirow}
\usepackage{caption}
\usepackage{subcaption}
\usepackage{xcolor}
\usepackage{slashbox}
\usepackage[font=scriptsize]{caption}

\newcommand\blfootnote[1]{%
  \begingroup
  \renewcommand\thefootnote{}\footnote{#1}%
  \addtocounter{footnote}{-1}%
  \endgroup
}

\title{Speaker Verification Across Ages: Investigating Deep Speaker Embedding Sensitivity to Age Mismatch in Enrollment and Test  Speech}
\name{Vishwanath Pratap Singh$^1$, Md Sahidullah$^2$, Tomi Kinnunen$^1$}
\address{
  $^{1}$University of Eastern Finland, Finland\\
  $^2$Institute for Advancing Intelligence, TCG CREST, India}
\email{vsingh@uef.fi, sahidullahmd@gmail.com, tomi.kinnunen@uef.fi}

\begin{document}

\maketitle
 
\begin{abstract}
In this paper, we study the impact of the ageing on modern deep speaker embedding based automatic speaker verification (ASV) systems. We have selected two different datasets to examine ageing on the state-of-the-art ECAPA-TDNN system. The first dataset, used for addressing short-term ageing (up to 10 years time difference between enrollment and test) under uncontrolled conditions, is VoxCeleb. The second dataset, used for addressing long-term ageing effect (up to 40 years difference) of Finnish speakers under a more controlled setup, is Longitudinal Corpus of Finnish Spoken in Helsinki (LCFSH). Our study provides new insights into the impact of speaker ageing on modern ASV systems. Specifically, we establish a quantitative measure between ageing and ASV scores. Further, our research indicates that ageing affects female English speakers to a greater degree than male English speakers, while in the case of Finnish, it has a greater impact on male speakers than female speakers. 

\end{abstract}


\blfootnote{A GitHub repository containing all the metadata and scripts will be published after the Interspeech decision.} 
\section{Introduction}
\vspace{-0.1cm}

Automatic speaker verification (ASV) systems have become increasingly important in the field of voice recognition technology including their applications in security systems, banking, call centers, law enforcement, fraud prevention, criminal investigations, and healthcare. Modern deep learning based ASV systems~\cite{bai2021speaker} involve three phases: (1)~training of speaker embedding extractor; (2)~enrollment to create a reference model of the target speakers; and (3)~verification to validate a speaker's claimed identity based on the similarity between the enrollment and verification recordings (ASV score).

\begin{figure}[htb]
\begin{minipage}[b]{1.0\linewidth}
  \centering
  \centerline{\includegraphics[width=5cm, height=7cm, keepaspectratio]{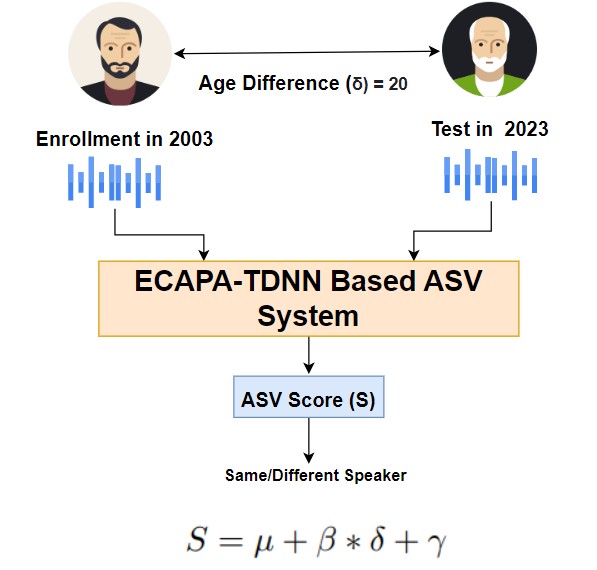}}

\end{minipage}
\vspace{-0.1cm}
\caption{We attempt  to explain the degraded ASV score (S) caused by the speaker ageing,  with respect to age difference ($\delta$)  across gender and different language using two  linear mixed effect model (LME) parameters namely, fixed effect due to the ageing ($\beta$) and random effect ($\gamma$) due to the various other factors across the sessions.} 
\label{fig:fig0}
\end{figure}

Similar to other biometrics, ASV systems are not perfect. It is evident that the accuracy of the ASV system depends strongly on how closely the enrollment and verification conditions are matched~\cite{m3,m2}. Over the years, extensive prior research works have addressed the effects of various mismatch factors between enrolment and verification phases, including background environment~\cite{mn1}, speaker-to-microphone distance~\cite{d1}, language~\cite{l1}, and speaking style~\cite{m3, rosa1, rosa2}. In this study, we focus on another fundamental mismatch factor that has received somewhat less attention: \emph{speaker ageing} due to the time lapse between the enrollment and verification recordings. 

\begin{table*}[t]
\scriptsize
 \caption{Age and Gender Enriched VoxCeleb1 and Longitudinal Corpus of Finnish Spoken in Helsinki (LCFSH) Details}
 \vspace{-0.1cm}
  \label{tab:age_vox1}
  \centering
  \begin{tabular}{|c | c|c |  c | c |c|c|c|} 
\hline
 \multicolumn{1}{|c|}{} &\multicolumn{3}{c|}{Male }  &\multicolumn{3}{c|}{Female }\\
\cline{2-7}
   Datasets & \# Speakers & \# Sessions& \# Utterances  & \# Speakers & \# Sessions& \# Utterances \\  
 \hline\hline
 VoxCeleb1 Official \cite{vox1} &690&12704&90450&561&9792&63066\\
 \hline
 VoxCeleb1 Enriched \cite{agen} &678&12476&88873 &536&9357&60306\\
 \hline
 VoxCeleb1-Age-Enriched (USA) &371&6584&46760 &299&5081&32303\\
 \hline
 LCFSH \cite{hlc} &51&51&7740 &58&58&7734\\
 \hline
 \end{tabular}
\end{table*}

\begin{table*}[t]
\scriptsize
 \caption{ Verification Accuracy (\%EER)  Across Gender on VoxCeleb1-Age-Enriched (USA). Here, session condition \emph{Both} includes recordings from both same and different sessions in trial pairs, while session condition \emph{Different} excludes recordings from the same session in trial pairs.}
 \vspace{-0.1cm}
  \label{tab:vox1}
  \centering
  \begin{tabular}{|c | c|c |  c | c |c |c |c|c | c |  c | c |c |} 
\hline
 \multicolumn{1}{|c|}{} &\multicolumn{12}{c|}{Age Difference  Between Enrolment and Test Pairs (Years) } \\
\cline{2-13}
   Gender & 0 (Both) & 0 (Different) &1  & 2 & 3 & 4 & 5 & 6 & 7 & 8 & 9 & 10  \\       
 \hline \hline
 Male &1.25&1.60&2.30&2.18&2.31&2.38&2.98&2.84&2.66&3.07&3.63&3.49\\
 \hline
 Female &1.34&1.59&2.29&3.91&2.70&2.70&3.05&3.12&3.66&3.41&3.87 & 3.89\\
 \hline
 \end{tabular}
\end{table*}

\begin{table}[t]
\scriptsize
 \caption{Verification Accuracy (\%EER)  Across Gender on Longitudinal Corpus of Finnish Spoken in Helsinki (LCFSH) Dataset.}
 \vspace{-0.1cm}
  \label{tab:lcfsh}
  \centering
  \begin{tabular}{|c | c|c |  c |} 
\hline
 \multicolumn{1}{|c|}{} &\multicolumn{3}{c|}{Age Difference   (Years) } \\
\cline{2-4}
   Gender & 0  &20  & 40  \\  
 \hline\hline
 Male &5.12&18.68&51.61\\
 \hline
 Female &10.04&21.67&40.61\\
 \hline
 \end{tabular}
\end{table}



\begin{figure*}
     \centering
     \begin{subfigure}[b]{0.24\textwidth}
         \centering
         \includegraphics[width=\textwidth,height=2cm]{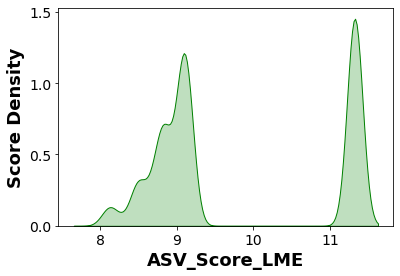}
         \caption{\tiny C-1 on Voxceleb1 Dataset}
         \label{fig:model-1-vox}
     \end{subfigure}
     \begin{subfigure}[b]{0.24\textwidth}
         \centering
         \includegraphics[width=\textwidth,height=2cm]{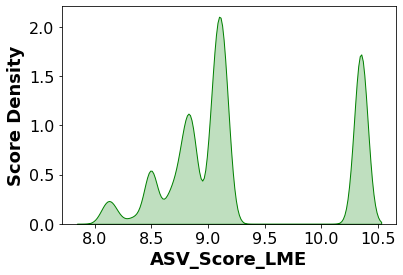}
         \caption{\tiny C-2 on Voxceleb1 Dataset}
         \label{fig:model-2-vox}
     \end{subfigure}
     \begin{subfigure}[b]{0.24\textwidth}
         \centering
         \includegraphics[width=\textwidth,height=2cm]{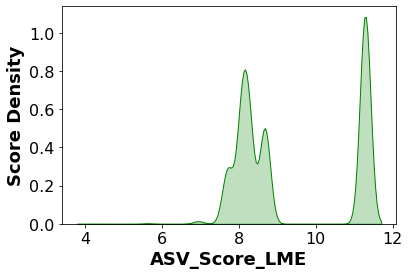}
         \caption{\tiny C-4 on Voxceleb1 Dataset}
         \label{fig:model-4-vox}
     \end{subfigure}
     \begin{subfigure}[b]{0.24\textwidth}
         \centering
         \includegraphics[width=\textwidth,height=2cm]{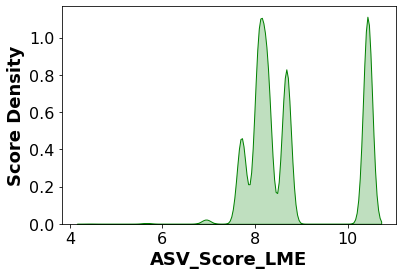}
         \caption{\tiny C-5 on Voxceleb1 Dataset}
         \label{fig:model-5-vox}
     \end{subfigure}
     \vspace{-0.1cm}
        \caption{Prediction of the ASV score density from LME on VoxCeleb1-Age-Enriched(USA) dataset under different conditions. The \textit{Statsmodels}  module in \textit{Python} enables score density prediction using the LME model \cite{slme}. Due to the space constraint, we have not plotted the non-target scores. }
        \label{fig:three graphs2}   
\end{figure*}

\begin{figure*}
     \centering
     \begin{subfigure}[b]{0.24\textwidth}
         \centering
         \includegraphics[width=\textwidth,height=2cm]{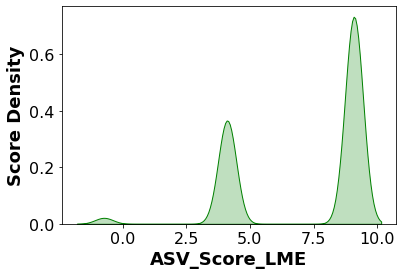}
         \caption{\tiny C-1 on LCFSH Dataset}
         \label{fig:model-1-hel}
     \end{subfigure}
     \begin{subfigure}[b]{0.24\textwidth}
         \centering
         \includegraphics[width=\textwidth,height=2cm]{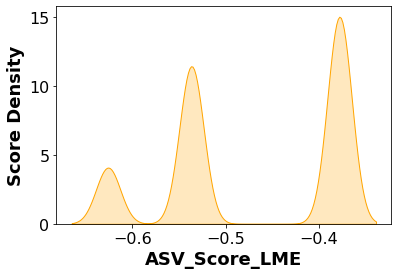}
         \caption{\tiny C-3 on LCFSH Dataset}
         \label{fig:model-3-hel}
     \end{subfigure}
     \begin{subfigure}[b]{0.24\textwidth}
         \centering
         \includegraphics[width=\textwidth,height=2cm]{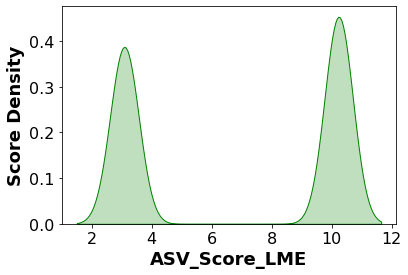}
         \caption{\tiny C-4 on LCFSH Dataset}
         \label{fig:model-4-hel}
     \end{subfigure}
     \begin{subfigure}[b]{0.24\textwidth}
         \centering
         \includegraphics[width=\textwidth,height=2cm]{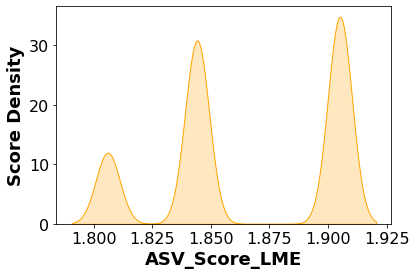}
         \caption{\tiny C-6 on LCFSH Dataset}
         \label{fig:model-6-hel}
     \end{subfigure}
     \vspace{-0.1cm}
        \caption{Prediction of the ASV score density from LME on LCFSH dataset under different conditions. The green plots are corresponding to target scores while orange plots are corresponding to non-target scores.}
        \label{fig:three graphs2}   
\end{figure*}

\begin{table*}[t]
\scriptsize
\caption{Various conditions for LME analysis and corresponding LME parameters. Here, session condition \emph{Both} includes recordings from both the same and different sessions in trial pairs, while session condition \emph{Different} includes recordings only from the different sessions in trial pairs. C1-C6 in Fig. 3 and Table 4 correspond each other. }
  \label{tab:lme}
  \centering
  \begin{tabular}{|c|c|c|c | c |  c | c |c |c |c|} 
 \cline{5-10}
 \multicolumn{4}{c|}{} &\multicolumn{3}{c|}{VoxCeleb1-Age-Enriched (USA)} & \multicolumn{3}{c|}{LCFSH}\\
\cline{1-10}
 Conditions  & Gender & Score & Session & Intercept & Fixed Effect  & Random Effect & Intercept & Fixed Effect & Random Effect \\  
 \hline\hline
C1 & Male & Target & Both & 9.74 & -0.164 & 0.493 &  8.664 & -0.246 & 0.029   \\
\hline
C2 &Male&Target& Different&9.453 &  -0.128 & 0.217  &  - & - & -   \\
\hline
C3 &Male&Non-target&   Different &-1.85 & -0.014
 & 	0.001 & 	-0.869  &	-0.002 & 	0.028 \\ 
 \hline
C4 &Female&Target&  Both&9.933 & -0.399     
 & 0.924 &	8.687 & -0.183 & 0.036 \\
 \hline
C5 &Female&Target&Different   &9.649 & -0.357
 & 0.669 &  	-	& - & - \\ 
 \hline
C6 &Female&Non-target& Different &-1.896 & -0.014
 &	0.001 &  	1.471	&  -0.001 & 0.005 \\ 
 \hline
 \end{tabular}
\end{table*}
\vspace{-0.1cm}

ASV systems utilize distinctive acoustic characteristics of an individual's voice to authenticate his or her identity. However, the impact of ageing on the human voice may alter these acoustic characteristics, which in turn can impact the performance of speaker verification systems. 
Specifically, as a person ages, their vocal tract, vocal folds, and related organs undergo changes that result in variations in measurable acoustic parameters, including pitch, formants~\cite{age3}, and loudness~\cite{age2}. Furthermore, the impact of the speaker ageing 
may depend substantially on speaker's gender and articulatory behavior~\cite{ageg}. Hence, a holistic study of ageing impact on the speaker verification systems should also consider the factors such as gender and language.

Despite having contributed to improved understanding of voice ageing and its impact on ASV systems, the prior studies have a number of limitations. In particular, the existing studies are limited by smallish sample sizes, controlled environments~\cite{ag1,ag2}, short-term ageing~\cite{ag3}, limited number of age groups and sessions~\cite{fage}. In addition, they have focused on classical GMM-UBM \cite{asv-gmm-ubm} and \emph{i-vector} based ASV systems \cite{ag1,ag2,score_cali}. In this paper, we have studied the impact of ageing on ASV scores across gender, datasets, languages, and different session conditions with ECAPA-TDNN \cite{ecapa} based ASV system.


The selection of appropriate datasets is vital for addressing the shortcomings of prior studies on age-related effects. In this study, we consider two datasets that are vastly different. The first dataset is VoxCeleb \cite{vox1,vox2}, a commonly used dataset to benchmark ASV systems. In order to examine the impact of ageing on English speech, we have restricted our analysis to speakers from the USA. 
The original VoxCeleb dataset, however, does not contain information about the age of speakers. In this study, we rely on the estimated age information in VoxCeleb provided in~\cite{agen}.

Our second dataset is \emph{Longitudinal Corpus of Finnish Spoken in Helsinki} (LCFSH) \cite{hlc}, collected between 1970 to 2013 with consistent intervals of approximately 20 years.
Unlike the VoxCeleb dataset which includes a mixture of different languages, accents, and recording conditions, the LCFSH dataset is exclusively in the Finnish language. Moreover, this is a controlled dataset with no uncertainty of the speaker's age. Hence, it is a good choice for studying the \emph{long-term} ageing effect on languages besides English.


Moreover, the impact of ageing on the ASV systems may have a combination of \emph{fixed effects} due to the age difference as well as \emph{random effects} due to various other factors across sessions. Hence, it is equally essential to consider utilizing appropriate analysis methods, such as \emph{linear mixed effect} (LME)~\cite{blme} models, which enable jointly modeling the random and fixed effects in ASV scores. LMEs are a type of statistical model used to analyze \emph{grouped data} using regression techniques.

To summarize, our study has the following main contributions:
\begin{itemize}
    \item Unlike most prior studies that have analyzed the ageing effect using a single language only, our study includes two very different languages and datasets namely, VoxCeleb and LCFSH. 
    
    \item Studies are conducted separately for male and female speakers to understand the impact of ageing across the gender.
    
    \item  As opposed to the previous studies that rely mean of scores, we have conducted extensive experiments using the linear mixed effect (LME) models to analyze the fixed and random effect on the ASV scores due to the ageing and other variations, such as recording setup and background noise.
    
    \item Unlike the existing studies which use traditional GMM-UBM or i-vector system, we perform all the experiments with state-of-the-art ECAPA-TDNN \cite{ecapa} based ASV systems.
    
\end{itemize} 

\begin{figure}[htb]
     \centering
     \begin{subfigure}[b]{0.21\textwidth}
         \centering
         \includegraphics[width=\textwidth, height=1.8cm]{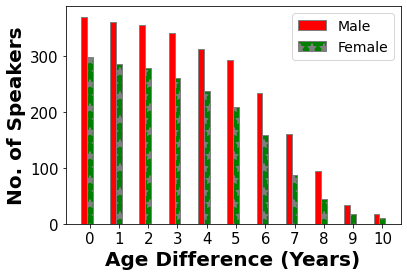}
         \caption{\tiny VoxCeleb1-Age-Enriched (USA)}
         \label{fig:figb-a}
     \end{subfigure}
     \begin{subfigure}[b]{0.21\textwidth}
         \centering
         \includegraphics[width=\textwidth,height=1.8cm]{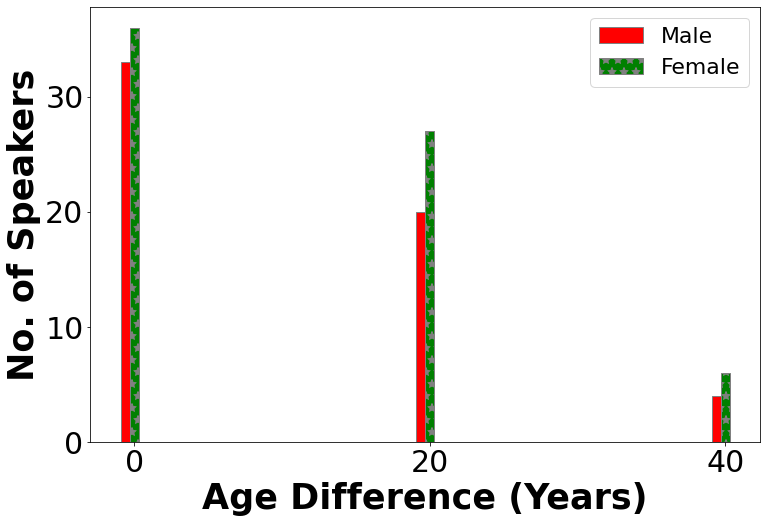}
         \caption{\tiny LCFSH}
         \label{fig:figb-b}
     \end{subfigure}
     \vspace{-0.1cm}
        \caption{Distribution of speakers with respect to age difference in (a.) VoxCeleb1-Age-Enriched (USA), and (b.) LCFSH dataset. }
        \label{fig:figb}   
\end{figure}
\vspace{-0.2 cm}
\section{Related Works}
\label{sec:relatedwork}
Prior research examining the effects of the ageing on ASV can be divided broadly into two main categories: (1)~investigating change in ASV scores; and (2)~analyzing and mitigating the impact of ageing on ASV accuracy. We briefly review both kinds of work below. 

\subsection{Analysing the Impact of Ageing on ASV Score}
In~\cite{ag1,ag2}, authors created a new dataset for studying ageing, comprising 9 male and 9 female speakers drawn from the British Broadcasting Corporation (BBC) and $Raidi\Grave{o}$ $Teilif\Grave{i}s$ $\Grave{E}ireann$ (RTE) broadcasts. The main findings from these two studies are: (1)~the effect of the ageing is more pronounced on the target (same-speaker) trial scores compared to the non-target (different speaker) trial scores; and (2)~the mean target score gets lowered due to the ageing, yielding increased equal error rates (EERs). Moreover, this study was conducted in controlled recording environments with a small sample size of speakers with classical GMM-UBM~\cite{asv-gmm-ubm} based ASV systems. 
In another work~\cite{ag3}, the study of short-term ageing on the GMM-UBM based ASV system was conducted with age differences up to three years. The authors concluded that the impact on the ASV due to the short-term ageing (0-3 years) was insignificant compared to the session variability. Similar to previous studies \cite{ag1,ag2}, the analysis is conducted using Multi-Session Audio Research
Project (MARP) corpus with a limited number of speakers, sessions, and in a gender independent manner. In \cite{fage}, only two age groups, one between 20-30 years of age (`young') and another between 40-50 years of age (`old'), were included for studying the effect of absolute age on factor analysis for speaker recognition. 

\subsection{Analyzing and Mitigating the Impact of Ageing on ASV Accuracy}
Besides the above studies, a number of studies have also addressed the impact of ageing on verification accuracy and proposed improvements to ASV systems in this regard. The impact of age difference between target and non-target speakers on speaker recognition performance has been investigated separately for male and female speakers~\cite{dodd}. In another study~\cite{score_cali}, the authors proposed the score calibration to mitigate the impact of the ageing on a i-vector based ASV system. Recently, an approach for \emph{age-invariant speaker embedding} was studied on the VoxCeleb dataset \cite{cage}. The primary focus of this study was to assess the improvement in the EER metric. 


To summarize, prior works on the impact of ageing on ASV systems have typically used small sample sizes. These studies have utilized conventional GMM-UBM and i-vector-based ASV systems. Furthermore, previous studies have focused primarily on the English language, with limited attention to other languages.



\vspace{-0.2 cm}
\section{Experimental Setup}

\subsection{Datasets}

In order to train the speaker embedding extractor, we have utilized the development portion of the VoxCeleb2 dataset~\cite{vox2}, which consists of 5,994 speakers. We randomly split the data into a 90:10 ratio for training and validation. Further, five additional samples for each utterance are generated using 2-fold time domain SpecAugment~\cite{specaugment} 
and 3-fold augmentations with combinations of publicly available MUSAN
dataset (babble-noise)~\cite{musan} and the room impulse response (RIR) dataset (reverb)~\cite{rir}. 

In the first part of the analysis, we have used the VoxCeleb1 dataset~\cite{vox1} for studying the impact of the ageing. We have utilized the gender, geography (country), and age metadata from \cite{agen}. To minimize the influence of geographic location, accent, and language differences we specifically chose to include only those sessions from VoxCeleb1 featuring speakers from the USA, the most represented country in the VoxCeleb1 dataset. We refer to this gender and age enriched dataset as \emph{VoxCeleb1-Age-Enriched (USA)} in the rest of the paper. Distributions of the number of speakers and sessions in the different evaluation protocols of \emph{VoxCeleb1-Age-Enriched(USA)} are presented in Table~\ref{tab:age_vox1}. Further,  Fig. \ref{fig:figb} illustrates the distribution of speakers in the VoxCeleb1-Age-Enriched (USA) dataset, categorized by gender.

In the second part, we have conducted experiments with the Longitudinal Corpus of Finnish Spoken in Helsinki (LCFSH)~\cite{hlc} dataset to study the impact the ageing in Finnish. The LCFSH dataset is collected  through interviews with the native residents  of Helsinki with varying social backgrounds. The dataset contains a total of 154 speakers of which 56 speakers have sessions with an age difference of approximately 20 years, and 13 speakers have sessions with an age difference of approximately 40 years. Number of speaker for each gender and age difference are illustrated in Fig.~\ref{fig:figb}. Number of speakers, sessions, and utterances available after the pre-processing are shown in Table \ref{tab:age_vox1}.
\vspace{-0.2 cm}
\subsection{Speaker Verification System}
\vspace{-0.1 cm}
We conduct all the model training and evaluation experiments using the SpeechBrain toolkit~\cite{sb}. Specifically, we use the ECAPA-TDNN based ASV system that achieves state-of-the-art performance~\cite{ecapa}. We use cosine similarity measure followed by the score normalized using adaptive s-norm \cite{snorm-1, snorm-2}. 
The ECAPA-TDNN model trained on VoxCeleb2 achieves EER of $1.28\%$ and $1.95\%$ on the Vox1-O and Vox1-E datasets, respectively. 


\section{Methodology}
\vspace{-0.1cm}
We use the linear mixed effect (LME) \cite{blme} models to analyze the variation in the ASV score with respect to the age difference under different gender, language, and session conditions. We fit the following model to ASV score data under different conditions:
\begin{equation}
    S_{i} = \mu + \beta *  \delta_{i} + \gamma 
\end{equation}
where $S_{i}$ is ASV score of $i$-th trial, $\mu$ is intercept or mean of the ASV scores in a particular condition, $\delta_{i}$ is the age difference between enrollment and verification audio, $\beta$ is fixed effect coefficient that establishes a linear relationship between age difference ($\delta_{i}$) and ASV score $S_{i}$, and $\gamma$ is a random effect which is induced by other variabilities such as background noise, recording setup, etc. We use the \textit{Statsmodels}  module in \textit{Python} for the LME analysis \cite{slme}.

In the previous study~\cite{ageg}, it is evident that the ageing impacts male and female speakers differently. Hence, we have conducted the analysis using different LMEs across different gender conditions. Further, the effect of the ageing on the ASV scores varies significantly across target and non-target scores~\cite{ag1, ag2}. Therefore, we also conduct separate analyses for target and non-target trial conditions. Finally, target trials from the same session can result in higher levels of similarity, which can introduce variations in ASV scores and create confusion with variations caused by the ageing. For this reason, we conduct separate analysis for different session conditions as well. 

Table~\ref{tab:lme} presents a total of six separate cases based on different gender, trial, and session conditions. When selecting target trial pairs with an age difference of zero in the \emph{VoxCeleb1-Age-Enriched (USA)} dataset, the pairs may originate from the same session or different sessions. For this study, we have taken into account both scenarios, including cases where trial pairs from the same session are present or excluded when there is no age difference between them. We represents these two session conditions as \emph{Both} and  \emph{Different} in Table \ref{tab:vox1} and \ref{tab:lme}. On the other hand, LCFSH corpus contains a single session for each speaker and hence the second session condition \emph{Different} is not applicable for this corpus and consequently C-2 and C-5 are not studied in Table \ref{tab:lme}.

\vspace{-0.3 cm}
\section{Results and Discussions}
\vspace{-0.1cm}
\subsection{Analyzing the Impact of Ageing on VoxCeleb1-Age-Enriched (USA)}
\vspace{-0.1cm}
Figure \ref{fig:figb-a} displays the gender and age distribution of speakers in VoxCeleb1-Age-Enriched (USA), while Table \ref{tab:vox1} presents the corresponding EERs. We observe in Table \ref{tab:vox1} that trends of degradation in EERs with respect to the ageing are not consistent. However, when observing the EERs at three-year intervals (i.e., EERs for age gaps of 0, 3, 6, and 9 years), we observe a monotonic increase in EERs. One possible explanation could be that the within-speaker variability is overshadowing the very short-term (0-2 years) ageing mismatches. 

Further, changes in intercepts of C1 \textit{v.s.} C2 and C4 \textit{v.s.} C5 in Table \ref{tab:lme}, as well as the plots in Fig.~\ref{fig:model-1-vox} \textit{v.s.} Fig.~\ref{fig:model-2-vox} and Fig.~\ref{fig:model-4-vox} \textit{v.s.} Fig.~\ref{fig:model-5-vox} indicate that the predicted target LME score is in a higher range when target trials from the same session are present. This suggests that the reason for a high LME score could be attributed to the similarity of the recording conditions or environment, particularly when trial pairs are chosen from the same session.

We also observe from the LME parameters in Table~\ref{tab:lme} that the fixed effect coefficients are negative in all the conditions (C1-6). This indicates both male and female speakers have a negative impact on ASV scores due to the ageing. In other words, as a consequence of aging, the similarities between embeddings decrease for all types of trials across different conditions. Further, we observe that non-target scores (C3 and C6) have low fixed effect coefficients which are considerably smaller in magnitude than that of target scores (C1, C2 and C4, C5). This indicates that the target scores are more sensitive to the speaker ageing compared non-target scores.

Additionally, by examining the parameters of C1 and C4 on the VoxCeleb1-Age-Enriched (USA) dataset in Table~\ref{tab:lme}, we observe that the fixed effect coefficient of C4 is comparatively greater than that of C1. This suggests that ageing might have a stronger influence on female target scores than on male target scores in the English language.

\vspace{-0.3 cm}
\subsection{Analyzing the Impact of Ageing on LCFSH}
\vspace{-0.1cm}
The EERs reported in Table~\ref{tab:lcfsh} for the LCFSH dataset is significantly higher than the  EERs reported in Table~\ref{tab:vox1} for VoxCeleb1-Age-Enriched (USA). This might be due to the domain mismatch between train and evaluation conditions between VoxCeleb and LCFSH induced by the differences in language, recording set-up, background environment, etc. Moreover, the degradation in EER within the LCFSH dataset for different ageing conditions is consistent and well-evident, likely due to age-groups with larger age gaps.

Similar to VoxCeleb1-Age-Enriched (USA), all fixed effect coefficients (C1, C3, C4, and C6) are negative in the LCFSH dataset. The fixed effect coefficients for target scores (C1 and C4) are notably larger in magnitude than those for non-target scores, indicating that, like in English, Finnish target scores are more sensitive to ageing than non-target scores. Furthermore, we observe in Fig.~\ref{fig:model-3-hel} and Fig.~\ref{fig:model-6-hel} that the x-axes for non-target scores are limited to a narrow range, while the target scores in Fig.~\ref{fig:model-1-hel} and  Fig.~\ref{fig:model-4-hel} are distributed more widely. This further confirms that target scores are more impacted due to ageing.

Unlike the results from VoxCeleb1-Age-Enriched (USA), the fixed effect coefficient for C4 is smaller than that for C1 in the LCFSH dataset. This suggests that, in Finnish, female target scores are less impacted than male target scores. Hence, we conclude that ageing impacts male and female speakers differently and the impact is not consistent across the languages. Furthermore, overall fixed effect coefficients of LME models trained on LCFSH dataset are smaller in magnitude than the corresponding LME models trained on VoxCeleb1-Age-Enriched (USA) dataset. This indicates that the ageing has less impact on Finnish compared to the English. This might be due to the differences in articulation styles~\cite{ageg} between the two languages.

\vspace{-0.2 cm}
\section{Conclusions}
\vspace{-0.2cm}
We examined the impact of ageing on the modern ECAPA-TDNN based speaker verification system. Using two publicly available datasets, we investigated the impact of ageing using a linear mixed effect model for different genders, session conditions, and languages. Our study reveals that the ageing impacts male and female speakers differently and the impact is dependent on the language. Considering trials from different sessions, we found that target scores are more dependent on age differences than non-target scores. This study utilized the estimated age for the speakers of the VoxCeleb dataset where the actual age might be different. This calls for further investigation of the issue with more accurate age information for this large dataset. The proposed methodology can also be useful for assessing age-invariant ASV systems.
\vspace{-0.2 cm}
\section{Acknowledgment}
\vspace{-0.1cm}
This work was partially supported by Academy of Finland (Decision No. 349605, project ``SPEECHFAKES'')
\bibliographystyle{IEEEtran}
\bibliography{mybib.bib}
\end{document}